\documentclass[reprint,prd,superscriptaddress]{revtex4-1}

\usepackage{amsmath}
\usepackage{amssymb}
\usepackage{graphicx}
\usepackage{dsfont}
\usepackage{float}
\usepackage{braket}
\usepackage{slashed}
\usepackage{epstopdf}
\usepackage{hyperref}

\newcommand{\newparagraph}{\newline\indent}
\newcommand{\bs}[1]{\boldsymbol{#1}}

\begin{document}
\title{Astrophysical constraints on dark-matter $Q$-balls in the presence of baryon-violating operators}

\author{Eric Cotner}
\affiliation{Department of Physics and Astronomy, University of California, Los Angeles, CA, 90095-1547, USA}

\author{Alexander Kusenko}
\affiliation{Department of Physics and Astronomy, University of California, Los Angeles, CA, 90095-1547, USA}
\affiliation{Kavli Institute for the Physics and Mathematics of the Universe (WPI), University of Tokyo, Kashiwa, Chiba 277-8568, Japan}

\begin{abstract}
Supersymmetric extensions of the standard model predict the existence of non-topological solitons, $Q$-balls.  Assuming the standard cosmological history preceded by inflation, $Q$-balls can form in the early universe and can make up the dark matter. The relatively large masses of such dark-matter particles imply a low number density, making direct detection very challenging.  The strongest limits come from the existence of neutron stars because, if a baryonic $Q$-ball is captured by a neutron star, the $Q$-ball can absorb the baryon number releasing energy and eventually destroying a neutron star.   However, in the presence of baryon number violating higher-dimension operators, the growth of a $Q$-ball inside a neutron star is hampered once the $Q$-ball reaches a certain size.  We re-examine the limits and identify some classes of higher-dimensional operators for which supersymmetric $Q$-balls can account for dark matter.  The present limits leave a wide range of parameters available for dark matter in the form of supersymmetric $Q$-balls.  
\end{abstract}


\maketitle

\section{Introduction}
Supersymmetric (SUSY) extensions of the standard model predict a scalar potential with a large number of flat directions \cite{ghergetta96}.  Such potentials admit stable configurations, SUSY $Q$-balls \cite{kusenko97.1,kusenko97.2,dvali99}.  Even if the scale of supersymmetry breaking is well above the reach of the present collider experiments, the flat directions can exist at a high scale and can play an important role in cosmology. If inflation took place in the early universe, a scalar condensate can form along the flat directions, leading to matter--antimatter asymmetry\cite{affleck85,dine95,Dine:2003ax}. In general, this scalar condensate is unstable with respect to fragmentation into $Q$-balls \cite{kusenko97.2,Kusenko:1997vp,Enqvist:1997si,Kasuya:1999wu}, which can be entirely stable and can play the role of dark matter \cite{kusenko97.2,Dine:2003ax,Enqvist:2003gh,Hong:2016ict}. This scenario offers a common origin to ordinary matter and dark matter. 

Dark-matter $Q$-balls have relatively large masses, and, therefore, very small number densities.  A direct detection of such dark matter is extremely challenging \cite{Kusenko:1997vp,Kasuya:2015uka}.
These flat directions are only flat at tree level, and in general they are lifted by non-renormalizable terms in the potential coming from loop corrections and GUT or Planck-scale physics, taking the form of polynomials in the squark fields and their conjugates
\begin{gather}
V_\text{lifting} = \frac{g}{\Lambda^{n+m-4}} \phi^n (\phi^*)^m + \text{c.c.} \label{eq:LiftingPotential}
\end{gather}
suppressed by some energy scale $\Lambda \sim 10^{16} \text{ GeV}$. If $n \not= m$, then baryon number is no longer conserved, fulfilling one of the Sakharov conditions for baryogenesis \cite{sakharov67}.  The same operators will destabilize the $Q$-ball \cite{Kawasaki:2005xc} and allow it to decay via processes that do not conserve the baryon number. If supersymmetric $Q$-balls make up the main component of dark matter, limits on their lifetimes (namely $\tau \gtrsim H^{-1}$) restrict the set of operators in the lifting potential in order to prevent their decay on too short of a timescale. \newparagraph
However, one can set additional constraints on the types of operators in the lifting potential by examining the effects of a star infected with a $Q$-ball. A $Q$-ball composed of squarks in the presence of baryonic matter absorbs the net baryon number and radiates pions on its surface \cite{kusenko04}. For a main sequence star, a $Q$-ball should pass through with a negligible change in velocity, due to the relatively low density of the star, and high inertia of the $Q$-ball. A neutron star, however, has a high enough density of baryons that a collision with a $Q$-ball should slow it to a crawl, at which point it would sink to the center of the star and begin to consume it from the inside out \cite{kusenko98,kusenko05}. If the $Q$-ball is absolutely stable, it grows without bound as it absorbs more neutrons until either the neutron star is completely consumed, or the $Q$-ball collapses into a black hole, causing the neutron star to collapse. Either way, we find the star dies relatively quickly on cosmological timescales, on the order of $10^8$ years.\newparagraph
However, the baryon number violation at a high scale is both plausible and necessary for the Affleck-Dine baryogenesis to work.  In the presence of baryon-number violating operators, the growth of a $Q$-ball inside a neutron star may be stymied by the baryon number destruction in the $Q$-ball interior, which becomes important once the $Q$-ball VEV reaches a certain magnitude \cite{kusenko05,Kasuya:2014ofa}.  
In this paper, we will re-examine the astrophysical bounds taking into account the baryon number violating operators.  The paper is organized as follows: section \ref{sec:QballStates} provides a brief review of allowed $Q$-ball states, section \ref{sec:DecayRates} explains the machinery of calculating the decay rate of the $Q$-ball, section \ref{sec:Qball+NS} details the interaction of the $Q$-ball with a neutron star, and section \ref{sec:BaryonEvolution} explains the evolution of the baryon number within the $Q$-ball and star. Section \ref{sec:Limits} takes this analysis and applies limits to the class of baryon-violating operators.

\section{Stable $Q$-ball states} \label{sec:QballStates}
We begin with a review of the stable ground states of $Q$-balls. The minimum necessary ingredients are a complex scalar field $\phi$ with a U(1) symmetry unbroken at the origin $\phi = 0$. Given a theory of multiple scalar fields with the action
\begin{gather}
S = \int d^4x\, \left[ \partial_\mu \phi_i^\dagger \partial^\mu \phi_i + \frac{1}{2} \partial_\mu \chi_j \partial^\mu \chi_j - V(\phi_i, \chi_j) \right]
\end{gather}
We can perform a Legendre transformation to get the Hamiltonian density of the theory, which gives us a functional for the energy.
\begin{gather}
E = \int d^3x\, \mathcal{H} \\
\mathcal{H} = |\dot{\phi}_i|^2 + |\nabla \phi_i|^2 + \frac{1}{2} \dot{\chi}_j^2 + \frac{1}{2}(\nabla \chi_j)^2 + V(\phi_i, \chi_j)
\end{gather}
Explicitly adding a Lagrange multiplier $\omega_i \left(Q_i - i \int d^3x \left(\dot{\phi}_i^\dagger \phi_i - \phi_i^\dagger \dot{\phi}_i\right)\right)$ to enforce charge conservation, we get a modified energy functional
\begin{gather}
\mathcal{E} = \int d^3x\, \tilde{\mathcal{H}} + \omega_i Q_i \label{eq:EnergyFunctional} \\
\tilde{\mathcal{H}} = |\nabla \phi_i|^2 + \frac{1}{2} (\nabla \chi_j)^2 + \tilde{V}(\phi_i, \chi_j) \\
\tilde{V}(\phi_i, \chi_j) = V(\phi_i, \chi_j) - \omega_i^2 |\phi_i|^2
\end{gather}
where we have assumed time dependence $\phi_i = \phi_i(\bs{x}) e^{i \omega_i t}$ and $\chi_j = \chi_j(\bs{x})$. If for any value of $\phi,\chi \not= 0$ and $0 < \omega_i < m$ there exists a point where $\tilde{V} < 0$, then stable $Q$-ball states exist. Furthermore, we can postulate that the stable states will be spherically symmetric, so that they depend only on the radial coordinate $r$.

\subsection{Flat direction $Q$-balls} \label{ssec:FlatDirection}
Assuming $V(\phi) \approx M^4 \sim (1 \text{ TeV})^4$ far from the origin, the vev in the interior of the $Q$-ball is not well-localized in $\phi$-space and the thin wall approximation does not hold. 
Instead, one can consider a thick-wall variational {\em ansatz} $\phi = \phi_0 \exp\left (-(r/R)^2\right )$. While the $r\rightarrow0$ behavior may be better described by $\sin (\omega r)/\omega r$, the analysis of Ref.~\cite{Enqvist:1998en} shows that the exponential {\em ansatz} is in good overall agreement with a numerical solution. 
Evaluating Eq.~(\ref{eq:EnergyFunctional}) with the assumption that $\int d^3x\, V \approx 4\pi R^3 M^4/3$, extremizing with respect to $R$ and using Hamilton's equation of motion $\omega = \dot{\theta} = \partial \mathcal{E}/\partial Q$, we arrive at
\begin{gather}
\omega = \pm M \left[ 4 \pi \cdot 3^{3/2} /Q \right]^{1/4} \qquad
\phi_0 = M \left[\frac{8 Q}{3^{3/2} \pi^2}\right]^{1/4} \nonumber \\
R = \frac{1}{M} \left[\frac{3^{1/2} Q}{4\pi}\right]^{1/4} \qquad
E = M \left[ 4 \pi \cdot 3^{3/2} Q^3 \right]^{1/4} \label{eq:FlatDirection}
\end{gather}
We can see that these types of $Q$-balls are stable in the large $Q$ limit since $\omega < m$ for large charge (the critical charge is $Q_c = 12\sqrt{3} \pi (M/m)^4$ with $m$ the mass of the scalar at $\phi = 0$), and $E \propto Q^{3/4}$. \newparagraph
$Q$-balls of this type are common in supersymmetric theories where a flat direction develops in the scalar potential for the superpartners of the quarks and leptons \cite{kusenko98,dvali99}. The conserved U(1) charge in these cases are then lepton and/or baryon number and are referred to in the literature as L-balls and B-balls. In addition to being able to form stable solitons, the interior of these $Q$-balls can sometimes support lepton- or baryon-violating vacuua \cite{dvali99}, which may be exploited in theories of baryo- or leptogenesis. Theories with charged inflatons may also be able to support these types of $Q$-balls since inflaton potentials must be relatively ``flat" to satisfy the slow-roll conditions.

\subsection{Curved direction $Q$-balls}
\begin{figure}
\centering
\includegraphics[width=0.7\linewidth]{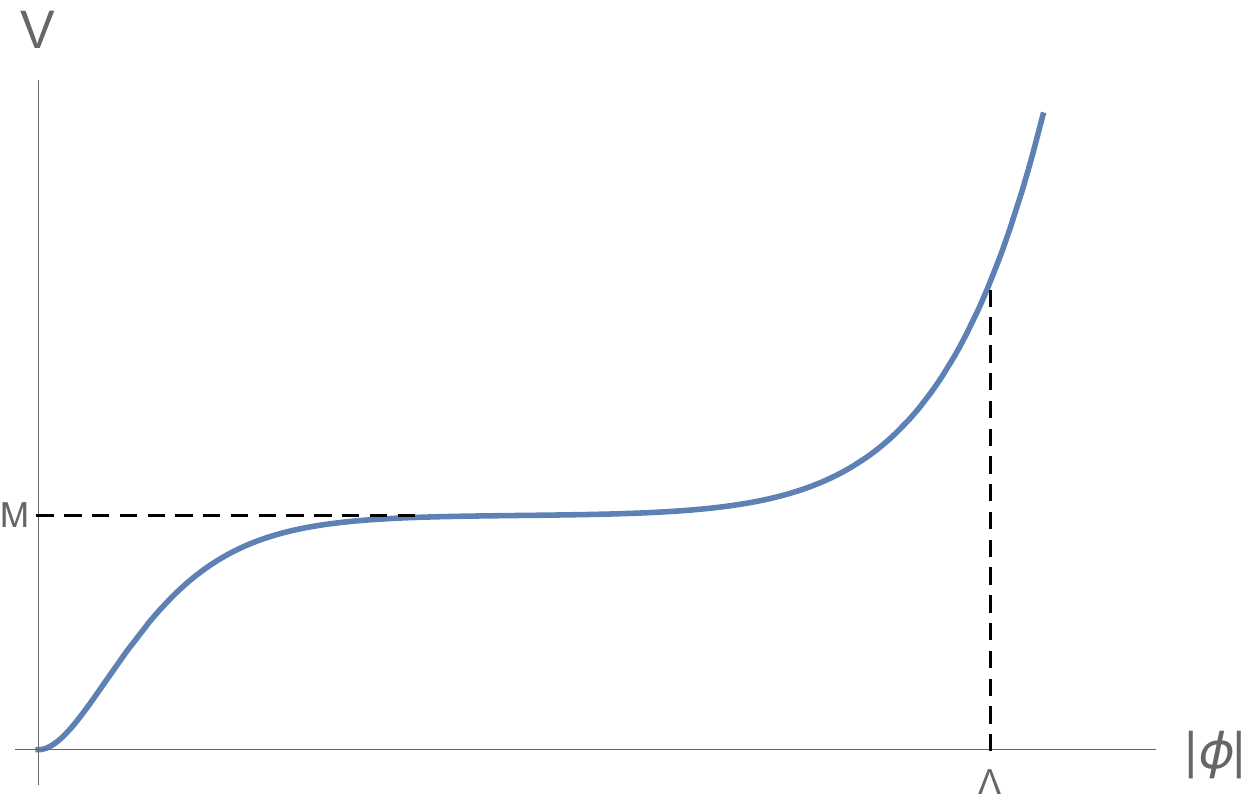}
\caption{Schematic scalar potential with a flat direction which is lifted by higher-dimension terms near $|\phi| \sim \Lambda$. Potentials of this form admit flat direction $Q$-balls which eventually grow into the curved direction type once the critical charge is surpassed.}
\label{fig:ScalarPotential}
\end{figure}

As the charge of a flat direction $Q$-ball grows, and the value of the scalar field vev $\phi_0$ slides to higher values, the corrections introduced by the lifting potential $V_\text{lifting}$ begin affecting the $Q$-ball (see figure \ref{fig:ScalarPotential}).
This happens when $\phi_0 \sim \Lambda$. If the lifting potential is of a form that respects the baryon number conservation, it can continue growing, albeit in a different manner. The vev hits a wall when it reaches its maximum at $\phi_0 = \Lambda$ and cannot climb any higher, so we can approximate the scalar potential near this point as
\begin{gather}
V(\phi) \approx M^4 + V_\text{lifting}^{\slashed{B}=0} = M^4 + \frac{2\,\text{Re}(g)}{\Lambda^{2n-4}} |\phi|^{2n}
\end{gather}
Since the vev is constrained to be near $\Lambda$, we can use the thin wall approximation. Substituting into equation \ref{eq:EnergyFunctional} and fixing $\phi_0 = \Lambda$, we vary with respect to $R$ to get
\begin{gather}
\omega = \pm \Lambda \sqrt{2\,\text{Re}(g) + (M/\Lambda)^4} \qquad \phi_0 = \Lambda \nonumber \\
R = \frac{1}{2\Lambda} \left[\frac{3Q}{\pi \sqrt{2\,\text{Re}(g) + (M/\Lambda)^4}}\right]^{1/3} \label{eq:CurvedDirection} \\
E = \Lambda \sqrt{2\,\text{Re}(g) + (M/\Lambda)^4} Q \nonumber
\end{gather}
Since we expect $M/\Lambda \ll 1$, we can neglect those terms under the square roots for simplicity. The critical charge at which point a flat direction $Q$-ball will become a curved direction $Q$-ball is $Q_c \approx 6.4(\Lambda/M)^4 \sim 10^{52}$. \newparagraph
If the lifting potential is not baryon-conserving, the U(1) symmetry is no longer respected and the $Q$-ball destabilizes, rapidly decaying until the lifting term is negligible and the $Q$-ball has reverted back to the flat direction type. Since curved direction $Q$-balls are necessarily more massive than the flat direction type (and their baryon consumption rate even faster), any limits obtained for flat direction $Q$-balls will also apply to the curved direction type, so we need only consider those belonging to the flat direction classification from now on.

\section{The decay rate} \label{sec:DecayRates}
We would now like to calculate the decay rate of the quanta of the $Q$-ball to other particles. The decay of $Q$-balls to neutrinos was first treated as an evaporation phenomenon due to the Pauli exclusion principle preventing decays in the interior of the $Q$-ball \cite{cohen86}. Bosons present no such obstacles, and therefore decays to scalar and vector particles can occur throughout the volume of the $Q$-ball, provided their mass is less than $\omega$. This may be difficult to achieve in general since most coupled scalar/gauge fields will get a mass term due to the nonzero expectation value in the $Q$-ball interior. However, the Nambu-Goldstone modes of the $Q$-ball field itself do not suffer this mass term, and decays to these modes can occur if the $U(1)$ symmetry is very slightly broken by the lifting potential. \newparagraph
Much work has been done calculating the decay and evaporation rates and energy spectra of $Q$-ball decays to fermions (both massless and massive) \cite{clark06,kawasaki13}. However, these previous studies did not treat decay of the condensate to bosons, and are related, but not relevant to the problem at hand. In this situation, we can utilize a simple method of calculating the decay rate that uses regular perturbation theory (with some extra steps).

\subsection{Mathematical background} \label{ssec:MathematicalBackground}
The probability for an initial state $\ket{\{\phi_i\}}$ to evolve into the final state $\ket{\{\phi_f\}}$ is given by $P = |\braket{\{\phi_f\}|\{\phi_i\}}|^2$. In the case of decays from a $Q$-ball, we are interested in the situation where the initial state is simply the scalar condensate describing the $Q$-ball: $\ket{\phi_c}$. Since the condensate is a persistent feature of the vacuum, the expectation value of the fields operator is simply the wave function: $\braket{\phi(x)} = \phi_c(x)$, a c-number function. $\phi_c(x)$ is the solution to the classical equations of motion in vacuum, which admit $Q$-ball solutions. Therefore, we can decompose the field operator into a classical and quantum part: $\phi = \phi_c + \hat{\phi}$ (we later employ a different decomposition in order to properly separate the field into its mass eigenstates, but it is conceptually similar to this one). \newparagraph
However, we are interested in how the $Q$-ball decays, so we must consider the state in which the scalar condensate is in the background of an interacting vacuum: $\ket{\Phi_c}$. The transition probability to any set of final state particles $\{\phi_f\}$ is then $P = |\braket{\Phi_c|\Phi_c \{\phi_f\}}|^2$. Using the single-particle expansion of the final particle states
\begin{gather}
\ket{\{\phi_f\}} = \int \prod_f \left(\frac{d^3p_f}{(2\pi)^3} \frac{\phi_f(\bs{p}_f)}{\sqrt{2E_f}}\right) \ket{\{\bs{p}_f\}}
\end{gather}
the transition probability is then
\begin{gather}
P = \left| \int \prod_f \left(\frac{d^3p_f}{(2\pi)^3} \frac{\phi_f(\bs{p}_f)}{\sqrt{2E_f}}\right) \braket{\Phi_c|\Phi_c \{\bs{p}_f\}} \right|^2
\end{gather}
Instead of using arbitrary wave functions as the final state, we can simply use the states of definite momentum, as is typically done, so that $\phi_f(\bs{p}_f) = (2\pi)^3 \delta^3(\bs{p}_f - \bs{p})/\sqrt{V}$. In this case, the differential transition probability is then
\begin{gather}
dP = \prod_f \left(\frac{d^3p_f}{(2\pi)^3} \frac{1}{2E_f}\right) |\braket{\Phi_c|\Phi_c \{\bs{p}_f\}}|^2 \label{eq:ProbabilityDensity}
\end{gather}
The matrix element $\mathcal{M} = \braket{\Phi_c|\Phi_c \{\bs{p}_f\}}$ can be computed perturbatively just as is normally done in QFT, except that we have to keep in mind the expansion of the scalar field operator $\phi = \phi_c + \hat{\phi}$. This leads to a bit of a complication, since working in the momentum space involves a Fourier transform of $\phi_c$, introducing an additional integral which consumes some of the delta functions that normally can be separated from the scattering amplitude $\mathcal{M}$. In addition, there is also no integral over the impact parameter since there are no collisions involved in this decay process. Depending on the number of interaction vertices in the process, we find the matrix element can be written schematically as
\begin{gather}
\mathcal{M} = \braket{\Phi_c|\Phi_c \{\bs{p}_f\}} = \mathcal{A}_n(\{\bs{p}_f\}) (2\pi) \delta(n\omega - \Sigma_f E_f)
\end{gather}
where $\omega$ is the $Q$-ball energy per particle (chemical potential), $n$ is the number of $Q$-ball quanta consumed by the decay (determined by the number of external legs attached to the condensate), and $\mathcal{A}$ is a ``reduced" matrix element. The delta function enforces global energy conservation, and although momentum is conserved at each vertex internal to the diagram, global momentum is not. This can be understood by the fact that the existence of the condensate breaks the spatial translation invariance of the vacuum, and therefore momentum is no longer a conserved quantity, the condensate instead absorbing the difference, similar to how a crystal lattice will absorb the recoil from a nuclear decay in the M\"ossbauer effect. \newparagraph
Now, one will find that equation \ref{eq:ProbabilityDensity} implicitly contains the square of a delta function, which is a little troubling. However, integration over the final state momenta will eat up one of the delta functions, leaving a $\delta(0)$, which is proportional to an infinite period of time $T = 2\pi \delta(0)$, in the sense that the limit of $T$ is this quantity, so that stripping this from the RHS gives us a probability per unit time per unit phase space; in other words, the differential decay rate
\begin{gather}
d\Gamma = \prod_f \left(\frac{d^3p_f}{(2\pi)^3} \frac{1}{2E_f}\right) |\mathcal{A}_n(\{\bs{p}_f\})|^2 (2\pi) \delta(n\omega - \Sigma_f E_f) \label{eq:DifferentialDecay}
\end{gather}
This method has wide applicability in calculating the decay of condensates and background fields, as the final state particles can be of either bosonic or fermionic type (the initial states can only be bosonic since fermions can't form condensates). The authors have also verified in the limit that the condensate wave function is that of a single zero-momentum particle $\phi_c \sim 1/\sqrt{V}$, the Fourier transform of which is a zero-momentum delta function, the decay rate reduces to that of a familiar single particle decay, as one would expect. The only drawback of this method is that it cannot handle decays that significantly alter the condensate wave function since $\phi_c$ would then be different in the initial and final states and it would not be appropriate to expand around. Thankfully, we will only be interested in decays involving $\Delta Q \lesssim 10$ from $Q$-balls with $Q \sim 10^{25}$, so the change in charge per decay is entirely negligible.

\subsection{Mass eigenstates and phonons} \label{ssec:MassEigenstates}
As briefly mentioned earlier, we would like to use a decomposition of the field operator that respects the mass eigenstates of the theory. For a theory with an unbroken U(1), a polar decomposition $\phi = \rho e^{i\theta}/\sqrt{2}$ shows that the scalar potential depends only on the radial field $\rho$. Therefore, this field is massive with the same mass as the original complex field: $m^2 |\phi|^2 = \frac{1}{2} m^2 \rho^2$. The potential is completely devoid of any terms containing $\theta$ however, due to the U(1) symmetry. This angular degree of freedom is therefore a massless Goldstone boson of the theory (inside the $Q$-ball it picks up a small mass due to the fact that has a minimum wavelength $\lambda \sim R$). Therefore we need a representation of the phonon operator that captures perturbations around the condensate while keeping the mass eigenstates separate. This leads us to consider the decomposition of the phonon field into a radial and angular part:
\begin{align}
\phi &= \frac{1}{\sqrt{2}} \rho e^{i\theta} = \frac{1}{\sqrt{2}} (\rho_c + \hat{\rho}) e^{i(\theta_c + \hat{\theta})} \nonumber \\
&\approx \phi_c + \frac{1}{\sqrt{2}} \hat{\rho} e^{i\omega t} + \frac{i}{\sqrt{2}} \hat{\psi} e^{i\omega t} + \cdots
\end{align}
where $\hat{\psi} \equiv \rho_c \hat{\theta}$, $\theta_c \equiv \omega t$, and the $\cdots$ refers to the higher-order terms in the Taylor expansion of the exponential. Although there is no way to invert the full relationship for $\hat{\rho}$ and $\hat{\psi}$ in terms of $\hat{\phi}$ and $\hat{\phi}^*$, the expansion to linear order can be inverted, and this gives us an approximate dictionary between the different phonon operators:
\begin{gather}
\hat{\phi} = \frac{1}{\sqrt{2}} (\hat{\rho} + i\hat{\psi}) e^{i\omega t} \quad \hat{\phi}^\dagger = \frac{1}{\sqrt{2}} (\hat{\rho} - i\hat{\psi}) e^{-i\omega t} \\
\hat{\rho} = \frac{1}{\sqrt{2}} \left( \hat{\phi}^\dagger e^{i\omega t} + \hat{\phi} e^{-i\omega t} \right) \quad \hat{\psi} = \frac{i}{\sqrt{2}} \left( \hat{\phi}^\dagger e^{i\omega t} - \hat{\phi} e^{-i\omega t} \right) \nonumber
\end{gather}
Unfortunately, we cannot simply substitute the above relationships into the Lagrangian because these are only correct to first order; we must expand around $\rho_c$ and $\theta_c$ in each term, then do a Taylor expansion in the exponential. \newparagraph
The $\hat{\phi}$ operator is complex, yet is not charged under the $U(1)$ of the theory inside the $Q$-ball because $\hat{\phi} \rightarrow \hat{\phi} e^{i\alpha}$ is not a symmetry of the Lagrangian unless $\phi_c = 0$ (in which case we are outside the $Q$-ball). Neither of the $\hat{\rho}$ or $\hat{\psi}$ is charged either, so a charged current cannot exist in the interior unless it is via bulk motion of, or interaction with, the condensate field $\phi_c$.

\subsection{Calculation of the matrix element}
\begin{figure}[h]
\centering
\includegraphics[width=0.8\linewidth]{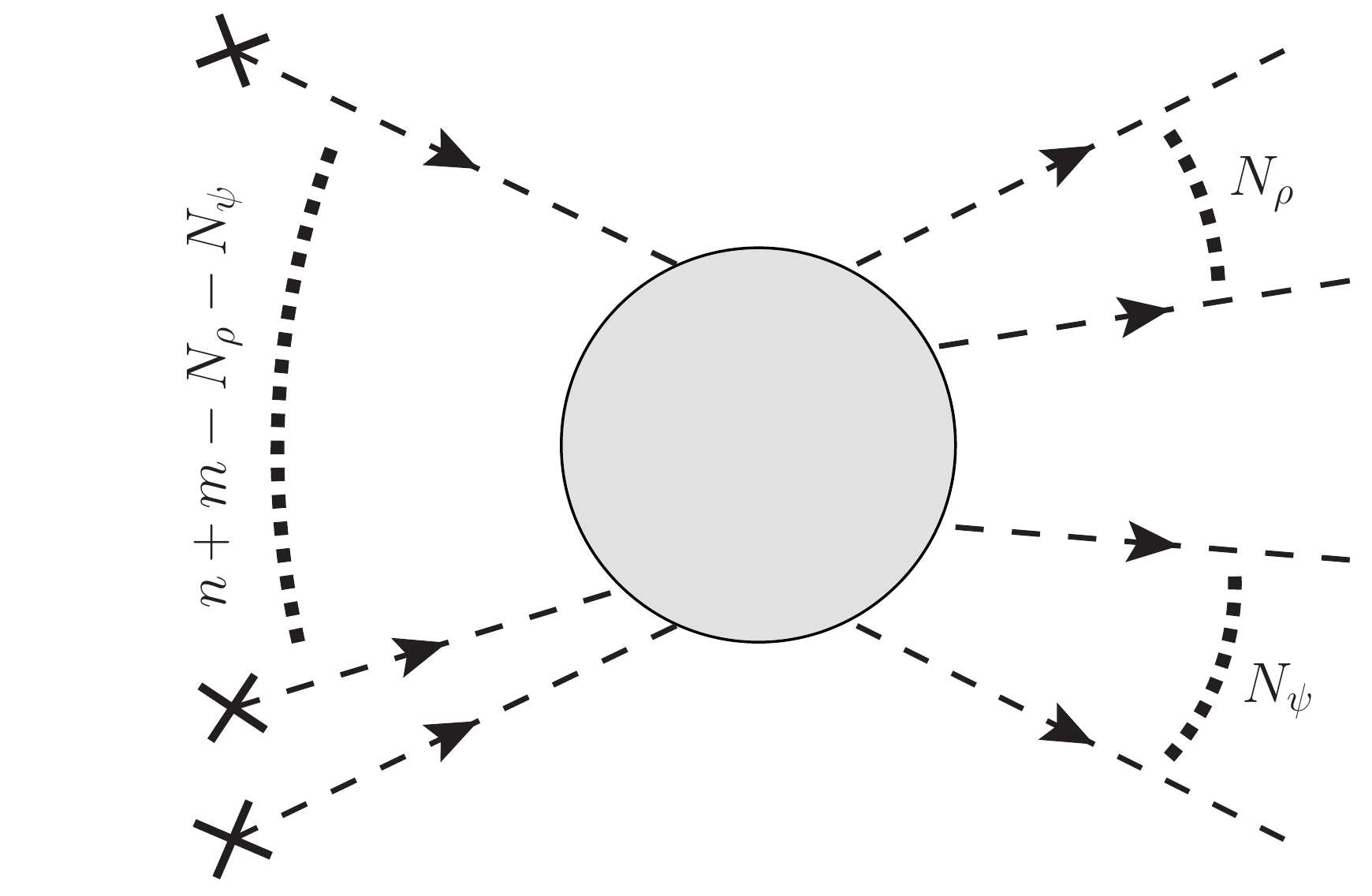}
\caption{Feynman diagram representation of the matrix element responsible for decay of the $Q$-ball into phonons. External lines on the left marked by a cross are interactions of the operator with the condensate $\phi_c$, whereas external lines on the right are the phonons produced from the decay. Arrows denote flow of momentum, not charge.}
\label{fig:CondensateFeynmanDiagram}
\end{figure}

We will now use the method of sections \ref{ssec:MathematicalBackground} and \ref{ssec:MassEigenstates} in order to derive the matrix element for decay of several $Q$-ball quanta to phonons within the $Q$-ball (the Feynman diagram representation of which is given by figure \ref{fig:CondensateFeynmanDiagram}).
We consider the lifting potential discussed earlier and expand it in polar form:
\begin{align}
\mathcal{L}_\text{lifting} &= -\frac{g}{\Lambda^{n+m-4}} \phi^n (\phi^\dagger)^m + \text{c.c.} \nonumber \\
&= -g_{nm} \left(\frac{\rho}{\sqrt{2}}\right)^{n+m} e^{i(n-m)\theta} + \text{c.c.}
\end{align}
where $g_{nm} \equiv g/\Lambda^{n+m-4}$. We now expand around the $Q$-ball condensate in the way prescribed above, giving us
\begin{widetext}
\begin{gather}
\mathcal{L}_\text{lifting} = - \frac{g_{nm}}{2^{(n+m)/2}} \sum_{j=0}^{n+m} \sum_{k=0}^{\infty} \binom{n+m}{j} \frac{i^k (n-m)^k}{k!} (\rho_c^{n+m-j-k} e^{i(n-m)\theta_c}) \hat{\rho}^j \hat{\psi}^k + \text{c.c.}
\end{gather}
Now we calculate the matrix element for the decay of the condensate to $N_\rho$ $\rho$'s and $N_\psi$ $\psi$'s:
\begin{align}
\mathcal{M} &= \frac{i}{2^{(n+m)/2}} \sum_{j,k} \binom{n+m}{j} \frac{(n-m)^k}{k!} \int d^4x\, \rho_c^{n+m-j-k} \left[i^k g_{nm} e^{i(n-m)\omega t} + \text{c.c.}\right] \nonumber \\
&\qquad\times \Braket{0| \hat{\rho}^j \hat{\psi}^k |p_1, \cdots, p_{N_\rho}, k_1, \cdots, k_{N_\psi}} \\
&= \frac{2\pi i}{2^{(n+m)/2}} C_{N_\rho N_\psi}^{nm} \int \left( \prod_{q=q_1}^{q_D} \frac{d^3q}{(2\pi)^3} \rho_c(\bs{q}) \right) (2\pi)^3 \delta^3\left(\bs{Q} - (\bs{P} + \bs{K})\right) \nonumber \\
&\qquad\times \left[ i^{N_\psi} g_{nm} \delta((n-m)\omega - (P^0 + K^0)) + (-i)^{N_\psi} g_{nm}^* \delta((m-n)\omega - (P^0 + K^0)) \right]
\end{align}
where $C_{jk}^{nm} \equiv j! \binom{n+m}{j} (n-m)^k$, $D \equiv n + m - N_\rho - N_\psi$ and $Q = \sum q$, $P = \sum p$, $K = \sum k$ are the sums of the various 4-momenta. We substitute in the $Q$-ball wave function to the Fourier transform $\rho_c(\bs{q}) = \sqrt{2} \phi_c(\bs{q}) = \sqrt{2} \pi^{3/2} R^3 \phi_0 e^{-q^2 R^2/4}$:
\begin{align}
\mathcal{M} &= \frac{i(2\pi)^{4-3D} 2^{D/2}}{2^{(n+m)/2}} C_{N_\rho N_\psi}^{nm} \left(\sqrt{2} \pi^{3/2} R^3 \phi_0 \right)^D \int \left( \prod_{q=q_1}^{q_D} d^3q\, e^{-\frac{R^2}{4} q^2} \right) \delta^3(\bs{Q} - (\bs{P} + \bs{K})) \nonumber \\
&\qquad\times \left[ i^{N_\psi} g_{nm} \delta((n-m)\omega - (P^0 + K^0)) + (-i)^{N_\psi} g_{nm}^* \delta((m-n)\omega - (P^0 + K^0)) \right]
\end{align}
Now, we use an interesting geometric argument to solve this integral. Since $d^3q = dq_1\, dq_2\, dq_3$ and $q^2 = q_1^2 + q_2^2 + q_3^2$, we observe that (besides the delta functions), the integral is hyper-spherically symmetric in the $3D$-dimensional $q$-space. The three delta functions each define a hyperplane in this space, the union of which is a $3(D-1)$-dimensional hypersurface which is a subspace of the larger $3D$-dimensional space. This hypersurface is displaced from the origin by the vector $\bs{v} = (\bs{P} + \bs{K})/\sqrt{D}$ (notice the hyperplane defined by $Q_i - (P_i + K_i) = 0$ has a unit normal vector of $\hat{n} = (1,1,\cdots,1)/\sqrt{D}$ and is displaced from the origin by a distance of $|P_i + K_i|/\sqrt{D}$). This integral therefore represents a spherically symmetric Gaussian integral over a $3(D-1)$-dimensional space offset from the origin by $\bs{v}$. We can therefore rotate our coordinate system so that $\bs{v}$ points in the new ``$\hat{z}$" direction and transform to a type of ``hypercylindrical coordinates": $(s,\phi,\theta_1,\cdots,\theta_{3(D-1)-2},x,y,z)$ where the coordinates $x$, $y$, and $z$ are Euclidean and a specification of $(x,y,z)=(0,0,v)$ constrains one to the hypersurface. Then, we simply perform the spherically symmetric integral over this surface:
\begin{gather}
\int \left( \prod_{q=q_1}^{q_D} d^3q\, e^{-\frac{R^2}{4} q^2} \right) \delta^3(\bs{Q} - (\bs{P} + \bs{K})) = \int d^{3(D-1)}s e^{-\frac{R^2}{4}(s^2 + v^2)} = \Omega_{3(D-1)-1} e^{-\frac{R^2}{4} v^2} \int_{0}^{\infty} ds\, s^{3(D-1)-1} e^{-\frac{R^2}{4} s^2}
\end{gather}
where $\Omega_{n-1} = 2\pi^{n/2}/\Gamma(n/2)$ is the solid angle of the $(n-1)$-sphere, and the remaining integral can be written in terms of a gamma function as well (it actually cancels with the one from $\Omega_{3(D-1)-1}$). After the dust has settled, the matrix element is found to be
\begin{align}
\mathcal{M} &= 2i \pi^{5/2} 2^{(n+m)/2 - N_\rho - N_\psi} C_{N_\rho N_\psi}^{nm} R^3 \phi_0^D e^{-\frac{R^2}{4D}(\bs{P} + \bs{K})^2} \nonumber \\
&\qquad\times \left[ i^{N_\psi} g_{nm} \delta((n-m)\omega - (P^0 + K^0)) + (-i)^{N_\psi} g_{nm}^* \delta((m-n)\omega - (P^0 + K^0)) \right]
\end{align}
The number of $Q$-ball quanta that decay in each event (and thereby amount of charge violation) can be read off from the delta function, and is $\Delta Q = |n-m|$, as expected. One important note is that since $m_\rho \gg \omega$, the condensate cannot decay to $\rho$'s unless $|n-m|\omega > m_\rho$, which requires the amount of charge violation to be very high. Decays to $\psi$'s might appear to proceed unimpeded, however, because they are massless. However, these phonons pick up a small mass from two different sources. First, as mentioned before, because the phonons are confined to the $Q$-ball, they are essentially standing waves with a maximum (Compton) wavelength of $\lambda \approx 2R$, which implies a minimum rest energy $m_\psi = 1/k = 1/4\pi R$. Since in a thick-wall $Q$-ball $\omega R = \sqrt{3}$, we have $m_\psi = \omega/4\pi\sqrt{3} \approx \omega/22$, which is small, but still a significant fraction of $\omega$! Second, the baryon-violating term itself introduces a small mass, which we can see by expanding to second order in $\hat{\psi}$:
\begin{gather}
\mathcal{L}_\text{lifting} \supset \frac{1}{2} \left[ \frac{(n-m)^2 \rho_c^{n+m-2}}{2^{(n+m)/2-1}} \left( \text{Im}(g_{nm}) \sin((m-n) \omega t) - \text{Re}(g_{nm}) \cos((m-n)\omega t) \right) \right] \hat{\psi}^2
\end{gather}
where the quantity in square brackets can be identified with $m_\psi^2$. Not only is this mass small in magnitude compared with the first contribution, but it also has harmonic time dependence, and therefore averages out to zero over timescales longer than about $|n-m|/\omega$. Thus, it is safe to assume that the mass from the Compton wavelength is the only mass that contributes.

\subsection{Calculation of the decay rate}
We now apply equation \ref{eq:DifferentialDecay} to calculate the decay rate, focusing on decays to only the Goldstone modes and setting $N_\rho = 0$ and $N \equiv N_\psi$. We take the squared amplitude (which can be simplified because the cross-terms are zero due to the conflicting delta functions), drop one of the delta functions, and integrate over the final state phase space:
\begin{gather}
\Gamma_{nm}^N = 4\pi^5 2^{n+m-2N} |g_{nm}|^2 (C_{0 N}^{nm})^2 R^6 \phi_0^{2D} I_{N}\left(|(n-m)|\omega,R/\sqrt{2D},m_\psi\right)
\end{gather}
where
\begin{gather}
I_N(\Omega,a,m) = \int \left( \prod_{p=p_1}^{p_N} \frac{d^3p}{(2\pi)^3} \frac{1}{2\sqrt{p^2 + m^2}} \right) e^{-a^2 \left(\sum \bs{p}\right)^2} \delta\left(\Omega - \sum \sqrt{p^2 + m^2}\right)
\end{gather}
For $N=1$ we can get an exact answer:
\begin{gather}
I_1(\Omega,a,m) = \frac{1}{(2\pi)^2} e^{-a^2(\Omega^2 - m^2)} \sqrt{\Omega^2 - m^2} \Theta(\Omega - m)
\end{gather}
However, using the relationships $m_\psi = \omega/4\pi\sqrt{3}$ and $R = \sqrt{3}/\omega$, we can reduce the integral to something even simpler:
\begin{gather}
I_{N}(|n-m|\omega,R/\sqrt{2D},m_\psi) = \int \left( \prod_{p=p_1}^{p_N} \frac{d^3p}{(2\pi)^3} \frac{1}{2\sqrt{p^2 + m_\psi^2}} \right) e^{-\frac{R^2}{2D} \left(\sum \bs{p}\right)^2} \delta\left(|(n-m)\omega| - \sum \sqrt{p^2 + m_\psi^2}\right) \nonumber \\
= \omega^{2N-1} \int \left( \prod_{p=p_1}^{p_N} \frac{d^3(p/\omega)}{(2\pi)^3} \frac{1}{2\sqrt{(p/\omega)^2 + (m_\psi/\omega)^2}} \right) e^{-\frac{24\pi^2}{D} \left(\sum (\bs{p}/\omega)\right)^2} \delta\left(|n-m| - \sum \sqrt{(p/\omega)^2 - (m_\psi/\omega)^2}\right)
\intertext{We then change coordinates to $\bs{\xi} \equiv \bs{p}/\omega$ and substitute the phonon mass so that $m_\psi/\omega \equiv \mu = 1/4\pi\sqrt{3}$:}
= \omega^{2N-1} \left[ \int \left( \prod_{\xi=\xi_1}^{\xi_N} \frac{d^3\xi}{(2\pi)^3} \frac{1}{2\sqrt{\xi^2 + \mu^2}} \right) e^{-\frac{24\pi^2}{n+m-N} \left(\sum \bs{\xi}\right)^2} \delta\left(|n-m| - \sum \sqrt{\xi^2 + \mu^2}\right) \right]
\end{gather}
where we will define the integral in square brackets as $J_{nm}^N$ (note $J_{nm}^N = J_{mn}^N$ and $J = 0$ if $N \ge n+m$ or $n=m$). Because $n,m$, and $N$ are integers and $J$ is a dimensionless number, we can simply tabulate all its possible values using numerical integration such as Monte Carlo. However, because of the delta function, we can't do MC until we integrate that out. We convert to spherical coordinates and separate the $N$th coordinate from the rest, then integrate over it to remove the delta function, leaving us with
\begin{align}
J_{nm}^N &= \frac{1}{(16\pi^3)^N} \prod_{\xi,\phi,\theta}^{}{}' \left( \int_{0}^{2\pi} d\phi \int_{0}^{\pi} d\theta\, \sin\theta \int_{0}^{\infty} \frac{\xi^2}{\sqrt{\xi^2 + \mu^2}} \right) \int_{0}^{2\pi} d\phi_N \int_{0}^{\pi} d\theta_N\, \sin\theta_N \xi_N(\{\xi\}) \nonumber \\
&\qquad\times e^{-\frac{24\pi^2}{n+m-N} \left( \sum_i^{'} \xi_i^2 + \sum_{i\not=j}^{'} \bs{\xi}_i \cdot \bs{\xi}_j + \xi_N^2(\{\xi\}) + 2\sum_{i}^{'} \xi_N(\{\xi\}) \xi_i (\sin\theta_N \sin\theta_i \cos(\phi_N - \phi_i) + \cos\theta_N \cos\theta_i) \right) } \label{eq:JMC}
\end{align}
where the primed sums/products mean we sum/multiply over all coordinates except the $N$th, and
\begin{gather}
\xi_N(\{\xi\}) \equiv \sqrt{\left(|n-m| - \sum{}' \sqrt{\xi^2 + \mu^2}\right)^2 - \mu^2} \\
\bs{\xi}_i \cdot \bs{\xi}_j = \xi_i \xi_j \left( \sin\theta_i \sin\theta_j \cos(\phi_i - \phi_j) + \cos\theta_i \cos\theta_j \right)
\end{gather}
Some sample values of $J_{nm}^N$ (I'll restrict to $n+m>4$ and $|n-m|=1$ for now) are:
\begin{align*}
N&=1:\qquad J_{23}^1 = 5.5 \times 10^{-28}, J_{34}^1 = 1.9 \times 10^{-19}, J_{45}^1 = 3.8 \times 10^{-15}, J_{56}^1 = 1.4 \times 10^{-12}, J_{67}^1 = 7.1 \times 10^{-11} \\
N&=2:\qquad J_{23}^2 = 1. \times 10^{-7}, J_{34}^2 = 3. \times 10^{-7}, J_{45}^2 = 5. \times 10^{-7}, J_{56}^2 = 8. \times 10^{-7}, J_{67}^2 1.2 \times 10^{-6} \\
N&=3:\qquad J_{23}^3 = 2. \times 10^{-11}, J_{34}^3 = 5. \times 10^{-10}, J_{45}^3 = 1. \times 10^{-9}, J_{56}^3 = 2. \times 10^{-9}, J_{67}^3 = 7. \times 10^{-9} \\
N&=4:\qquad J_{23}^4 = 8. \times 10^{-17}, J_{34}^4 = 6. \times 10^{-14}, J_{45}^4 = 9. \times 10^{-13}, J_{56}^4 = 5. \times 10^{-12}, J_{67}^4 = 2. \times 10^{-11} \\
N&=5:\qquad J_{34}^5 = 3. \times 10^{-19}, J_{45}^5 = 7. \times 10^{-17}, J_{56}^5 = 3. \times 10^{-15}, J_{67}^5 = 2. \times 10^{-14}
\end{align*}
Clearly, final states involving more phonons have a smaller amount of phase space volume. The exception is $N=1$, which gets extra suppression from the fact that any decay involving one final state particle does not conserve momentum. It should be noted that repeated evaluation of the Monte Carlo shows that the uncertainty in these answers is quite large; variation in the first digit is common, though the order of magnitude remains consistent over repeated evaluations. It turns out that this is not terribly important for computing the neutron star lifetimes; variations of $O(1)$ in $J$ translate to variations of $O(10^{-3})$ in the lifetimes. This is because the dependence of $\Gamma$ on $Q$ is most important. There is also a small imaginary part attached to some of these numbers which is not shown. This is from integrating over a region in phase space which is not kinematically allowed, and it does not contribute to the decay rate, so we can simply ignore it. \newparagraph
If the dimension of the lifting potential is extremely high ($n+m \rightarrow \infty$), then the exponential in the integrand becomes order unity, and we can reduce this even further by transforming to a dimensionless energy coordinate $\sigma = \sqrt{\xi^2 + \mu^2}$ and integrating out all the angles. The integral $J_{nm}^N$ approaches
\begin{gather}
J_{nm}^N \rightarrow \frac{1}{(2\pi)^{2N}} \left( \prod_{\sigma=\sigma_1}^{\sigma_N} \int_{\mu}^{|n-m|-(N-1)\mu} d\sigma\, \sqrt{\sigma^2 - \mu^2} \right) \delta\left(|n-m| - \sum \sigma \right)
\end{gather}
This can be calculated via Monte Carlo in a similar manner to equation \ref{eq:JMC}. Note that in this limit $J_{nm}^N$ only depends on $\Delta Q = |n-m|$ and $N$. Now, putting all of this together, we can express the decay rate to $N$ Goldstone bosons as
\begin{gather}
\Gamma_{nm}^N = 4\pi^5 2^{n+m-2N} |g_{nm}|^2 (N!)^2 (c_{0 N}^{nm})^2 R^6 \phi_0^{2D} \omega^{2N-1} J_{nm}^N
\end{gather}
Writing out $R$, $\phi_0$ and $\omega$ in terms of $Q$ (see equations \ref{eq:FlatDirection}) and lumping all the non-dimensional constants together,
\begin{gather}
\Gamma_{nm}^N = |g|^2 K_{nm}^N Q^{\frac{1}{4}(7+2(n+m-2N))} M \left(\frac{M}{\Lambda}\right)^{2(n+m)-8} \\
K_{nm}^N \equiv 2^{\frac{1}{2}(5(n+m-N)-3)} 3^{\frac{3}{8}(1-2(n+m-2N))} \pi^{\frac{13}{4}-(n+m-3N/2)} (n-m)^{2N} J_{nm}^N
\end{gather}
where $M$ is the mass scale associated with the potential energy density in the flat direction of the scalar potential ($V_0 = M^4$). We can now simply tabulate the $K_{nm}^N$ and have a semi-analytic expression for the decay rate that will be easy to use in the analysis of section \ref{sec:BaryonEvolution}.
\end{widetext}

\section{Interactions between $Q$-balls and neutron stars} \label{sec:Qball+NS}
We would now like to understand how a $Q$-ball interacts with its host star in order to determine the neutron consumption rate. As discussed in the work of one of us, Loveridge, and Shaposhnikov (KLS) \cite{kusenko05}, the transport mechanism of neutrons inside a neutron star is complicated and is not very well understood. The authors outline two different possible situations for neutron accretion, which we will summarize here for clarity.

\subsection{Surface conversion of neutron flux}
As a rough estimate, KLS assume the rate of neutron absorption is simply equal to the flux of neutrons moving across the surface of the $Q$-ball. In this scenario, the growth rate of the $Q$-ball is given by
\begin{align}
\dot{Q} &= b^{-1} 4\pi R^2 n_0 v = \frac{4 \cdot 3^{5/4} n_0 }{M^2 (4\pi)^{1/2}} Q^{1/2} \nonumber \\
&\approx (2 \times 10^{-8} \text{ GeV}) Q^{1/2}
\end{align}
where $b=1/3$ is the baryon number of a squark, $n_0 \approx 10^{15} \text{ g/cm}^3 = 4 \times 10^{-3} \text{ GeV}^{3}$ is the neutron number density at the center of the star, and $v \approx 1$ is the speed of the neutrons, assumed to be of the order of the speed of light. This estimate for the absorption rate is likely too high, as it does not take into account the pressure backreaction from the pions and antineutrons produced on the surface of the $Q$-ball.

\subsection{Hydrodynamic considerations due to pion production}
Using a couple different methods, KLS determine the pressure at the center of the star in hydrostatic equilibrium is approximately $P \approx (0.1 \text{ GeV})^4$. For light degrees of freedom such as pions, electrons and neutrinos, this implies a temperature of about $100 \text{ MeV}$ from the relation $P \approx g T^4/\pi^2$. This temperature cannot be maintained by thermal effects alone, but can be maintained by the pions produced on the surface of the $Q$-ball. The rate of pion loss to decay inside the star is given by
\begin{gather}
\dot{N}_\pi \approx 2\pi^{3/2} \sqrt{\frac{\lambda}{3\tau}} n_\pi(0) R^2
\end{gather}
where $\lambda \approx n_0^{-1/3}$ and $\tau \approx 10^8 \text{ GeV}^{-1}$ are the mean free path and neutral pion lifetime, respectively. They also assume $n_\pi(0) \approx n_0$ in order to maintain pressure. Each neutron only has enough mass and energy to supply about 4-5 pions, so the rate of neutron absorption is about that much lower, giving us
\begin{align}
\dot{Q} &= \frac{10 \pi^{3/2} n_0^{5/6} R^2}{b \sqrt{3\tau}} = \frac{5\pi n_0^{5/6}}{3^{1/4} b M^2 \tau^{1/2}} \nonumber \\
&\approx (10^{-11} \text{ GeV}) Q^{1/2}
\end{align}
This estimate is slightly lower than the raw neutron flux estimate and is a little more realistic. 

\section{Baryon number evolution in an infected neutron star} \label{sec:BaryonEvolution}
\begin{figure}
\centering
\includegraphics[width=1.0\linewidth]{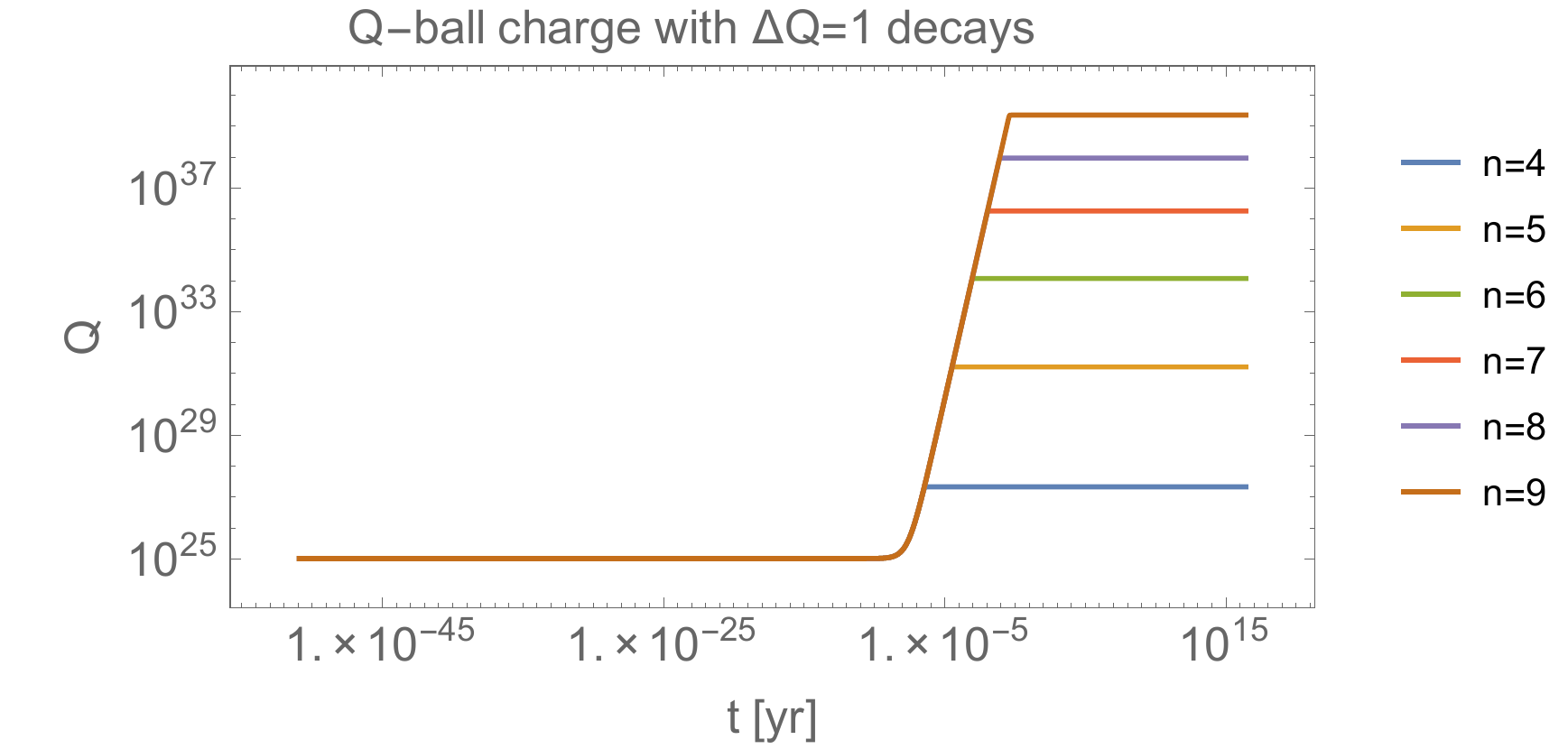}
\caption{Plot of the evolution of charge within the $Q$-ball at the center of a neutron star with decay channels attributed to various $\Delta Q = 1$ operators, indexed by $(n,m) = (n,n+1)$. The $Q$-ball very quickly equilibrates so that the rate of decay is equal to the rate of neutron consumption. Not shown are the contours for $n=2$ and $n=3$, which are ruled out because the corresponding operators would destabilize the $Q$-ball in free space on short timescales.}
\label{fig:QEvolution}
\end{figure}

\begin{figure}
\centering
\includegraphics[width=0.9\linewidth]{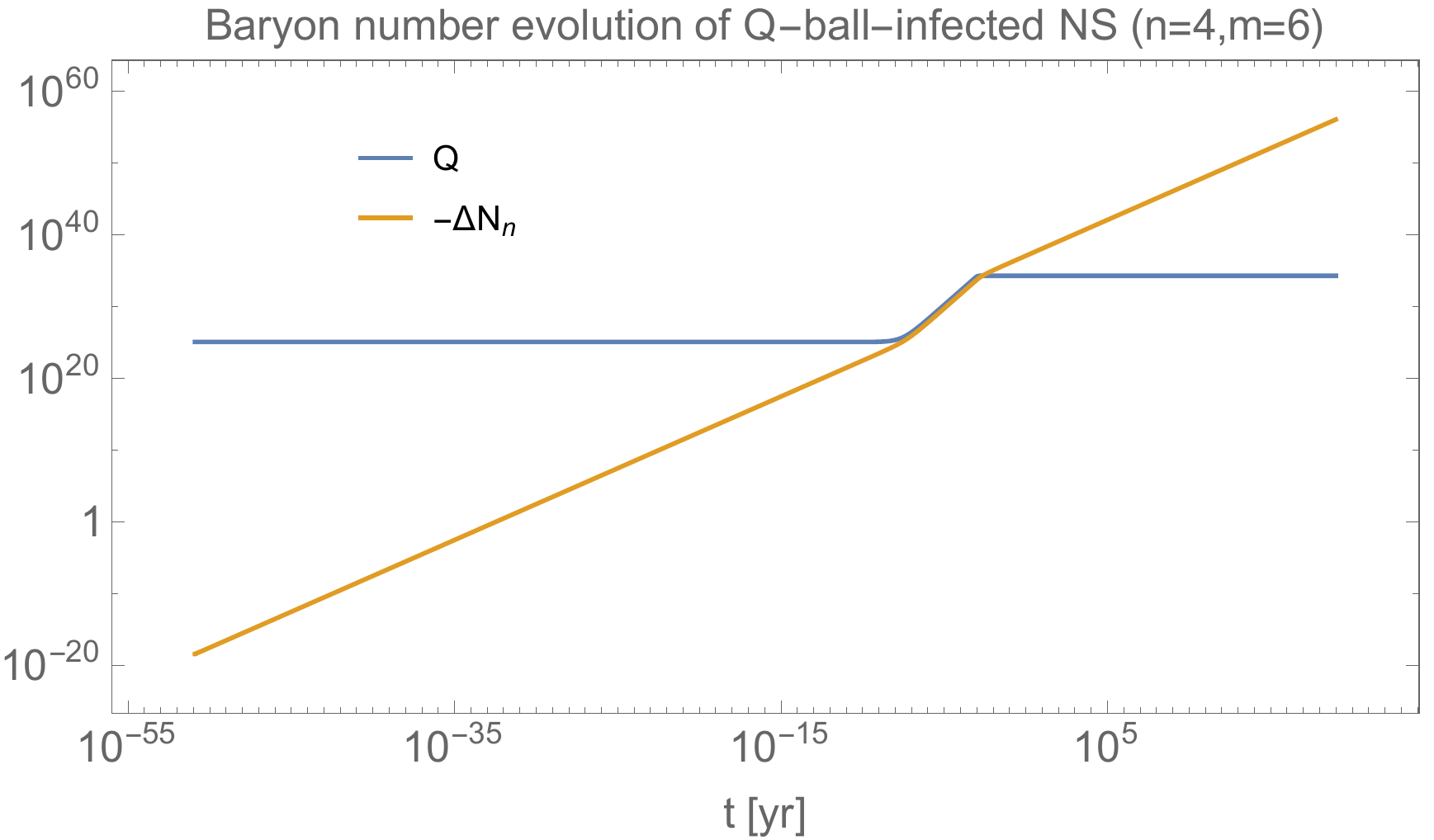}
\caption{Plot of the charge $Q$ contained within a $Q$-ball and the number of neutrons consumed by the $Q$-ball over the life of the neutron star. Once $-\Delta N_n = B_\text{NS} = 10^{57}$, integration is stopped and the star is gone. This specific example is for a $Q$-ball with decays mediated by a $(n,m)=(4,6)$ operator, resulting in a neutron star lifetime of $\tau_\text{NS} = 1 \times 10^{20}$ years.}
\label{fig:BaryonNumberEvolution}
\end{figure}

Now that have expressions for both the growth rate and decay rate of the $Q$-ball, we can set up a simple set of differential equations to model the evolution of the baryon number in both the $Q$-ball and the neutron star:
\begin{gather}
\dot{B}_Q = b \dot{Q} = -\dot{N}_n - b\, |n-m|\, \Gamma_{nm} \label{eq:QballEvolution} \\
\dot{B}_\text{NS} = \dot{N}_n = -(10^{-11} \text{ GeV}) Q^{1/2} \label{eq:NSEvolution}
\end{gather}
where $\Gamma_{nm} \equiv \sum_N \Gamma_{nm}^N$. Or, eliminating $N_n$ and assuming decays are dominated by a specific $N$ (usually either 1 or 2), we can put it in a more aesthetically pleasing form:
\begin{gather}
\dot{Q} \approx 3 \dot{N}_0 Q^{1/2} - \Gamma_0 Q^\alpha \label{eq:QEvolutionSimplified}
\end{gather}
where $\dot{N}_0 = 10^{-11} \text{ GeV}$, $\Gamma_0 = |g|^2 K_{nm}^N M (M/\Lambda)^{2(n+m)-8}$, and $\alpha = \frac{1}{4}(7+2(n+m-2N))$ ($\alpha > 1$ unless $N$ is some ridiculously high number, which is unlikely). The initial conditions for this system are $Q(0) = Q_0 \approx 10^{25}$ and $N_n(0) = B_\text{NS} = 10^{57}$, and the total number of neutrons absorbed by the $Q$-ball is given by integrating equation \ref{eq:NSEvolution}:
\begin{gather}
\Delta N_n(t) = -\int_{0}^{t} dt'\, \dot{N}_0 Q^{1/2} \label{eq:NeutronDepletion}
\end{gather}
We can see that equation \ref{eq:QEvolutionSimplified} has late-time attractor solutions, whereby setting $\dot{Q} = 0$, we solve for the equilibrium charge: $Q_\text{eq} = (3\dot{N}_0/\Gamma_0)^{\frac{1}{\alpha - 1/2}}$ (see figure \ref{fig:QEvolution}).
If this charge is reached relatively quickly compared to the total lifetime of the neutron star, then equation \ref{eq:NeutronDepletion} implies that the neutron depletion is linear in time, and the lifetime of the star is then
\begin{gather}
\tau_\text{NS} \approx \frac{B_\text{NS}}{\dot{N}_0} \left(\frac{3 \dot{N}_0}{\Gamma_0}\right)^\frac{1/2}{1/2 - \alpha}
\end{gather}
In free space, the evolution of the charge of a $Q$-ball is given by equation \ref{eq:QEvolutionSimplified} with $\dot{N}_0 = 0$, which can easily be solved for:
\begin{gather}
Q(t) \approx \left[(\alpha - 1) \left( \frac{Q_0^{1-\alpha}}{\alpha - 1} + \Gamma_0 t \right)\right]^\frac{1}{1-\alpha}
\end{gather}
The $Q$-ball lifetime can then be solved for by setting $Q(\tau_Q) = 1$, which gives us
\begin{gather}
\tau_Q \approx \frac{1 - Q_0^{1-\alpha}}{(\alpha - 1) \Gamma_0}
\end{gather}
If we want to be more exact and take into account decays from all channels (not just the dominant one), we can numerically solve for $\tau_\text{NS}$ and $\tau_Q$ by evolving equations \ref{eq:QballEvolution} and \ref{eq:NSEvolution} until $N_n = 0$ or $Q=1$, at which point either the neutron star has been consumed or the $Q$-ball has decayed, and we stop integration (a specific example is given in figure \ref{fig:BaryonNumberEvolution}).
This is how we will derive the limits in the next section.

\section{Limits on baryon-violating (and conserving) operators} \label{sec:Limits}
Using equations \ref{eq:QballEvolution} and \ref{eq:NSEvolution} and the algorithm prescribed in the previous section, we can tabulate the lifetimes of infected neutron stars and free $Q$-balls endowed with the lifting potential of equation \ref{eq:LiftingPotential}, indexed by the integers $n$ and $m$. We will find that baryon-violating terms are necessary if an infected neutron star is to survive to present day.

\subsection{From decay of $Q$-balls in free space} \label{ssec:QballLifetime}
We solve the baryon number evolution equations with $\dot{N}_0 = 0$ in order to model the decay of the $Q$-ball in free space. The results are plotted in figure \ref{fig:QballLifetimes} and tabulated in table \ref{tab:QballLifetimes} in the appendix.
\begin{figure}
\centering
\includegraphics[width=0.8\linewidth]{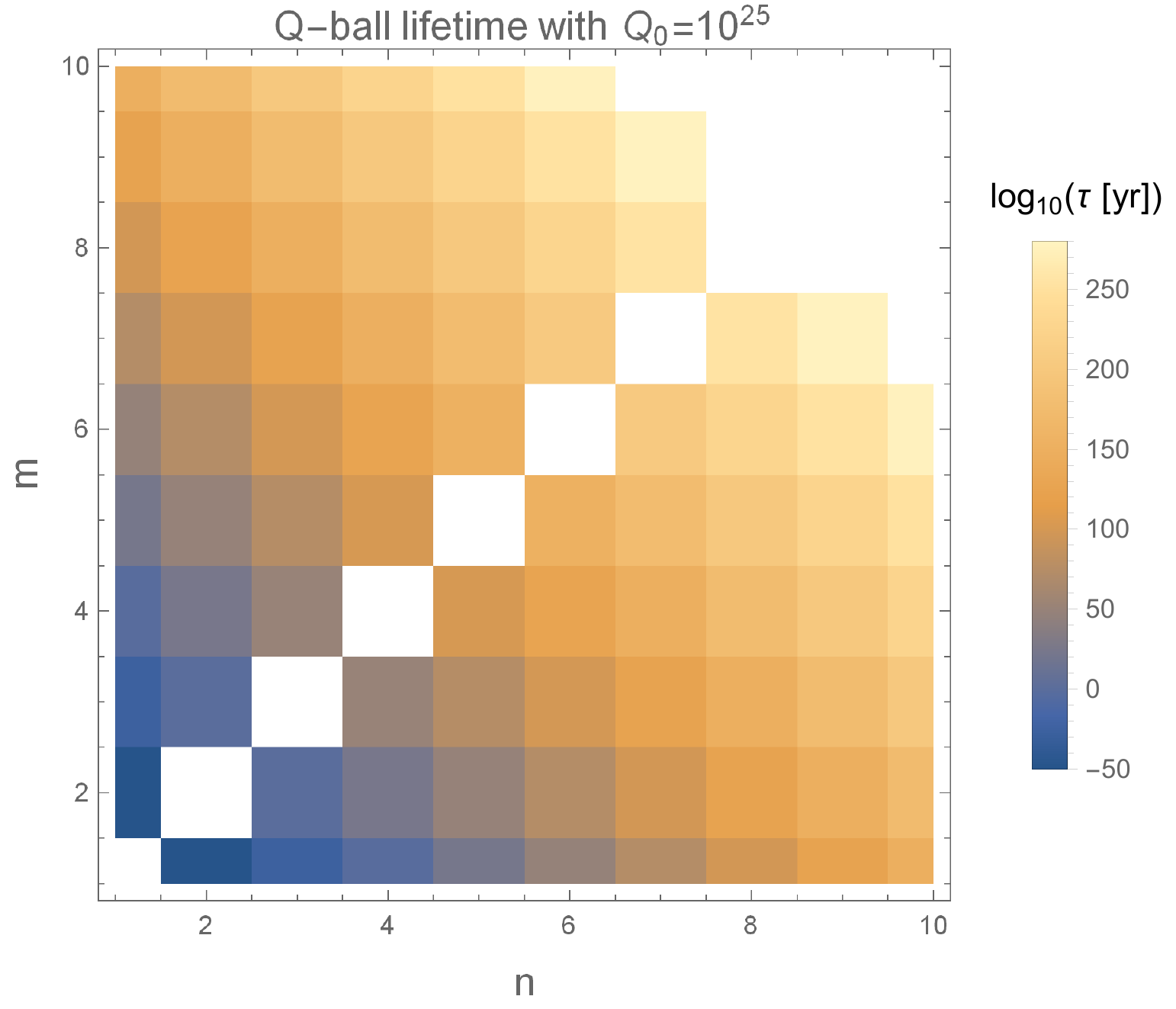}
\caption{Plot of $Q$-ball lifetimes with an initial charge of $Q_0 = 10^{25}$ as a function of various $n$, $m$ corresponding to the terms in the lifting potential. The diagonal $n=m$ is actually completely stable because decays are not permitted due to restoration of the $U(1)_B$ symmetry.}
\label{fig:QballLifetimes}
\end{figure}
The most striking feature is that for $n=m$, the $Q$-ball is completely stable because the Goldstone field does not appear in the potential. We can also see that in general, as the dimension of the operator increases, so does the lifetime of the $Q$-ball. In fact, all $Q$-balls with lifting potentials of dimension 5 or less are unstable and decay in a matter of hours or less, whereas those with dimension greater than 5 are stable on timescales much longer than the age of the Universe. This immediately rules out dark matter $Q$-balls with $n+m \le 5$. In the high-dimension limit ($n+m \rightarrow \infty$), we can calculate $J_{nm}^N$ using equation \ref{eq:JMC} and solve the baryon evolution equations again, though this doesn't lead to any interesting revelations; the $Q$-ball lifetime continues to increase as the dimension of the operator increases, and is pretty much independent of $\Delta Q$. The largest lifetime calculated (dim $=100$) was over $10^{2000}$ years!

\subsection{From lifetime of neutron stars} \label{ssec:NSLifetime}
\begin{figure}
\centering
\includegraphics[width=0.8\linewidth]{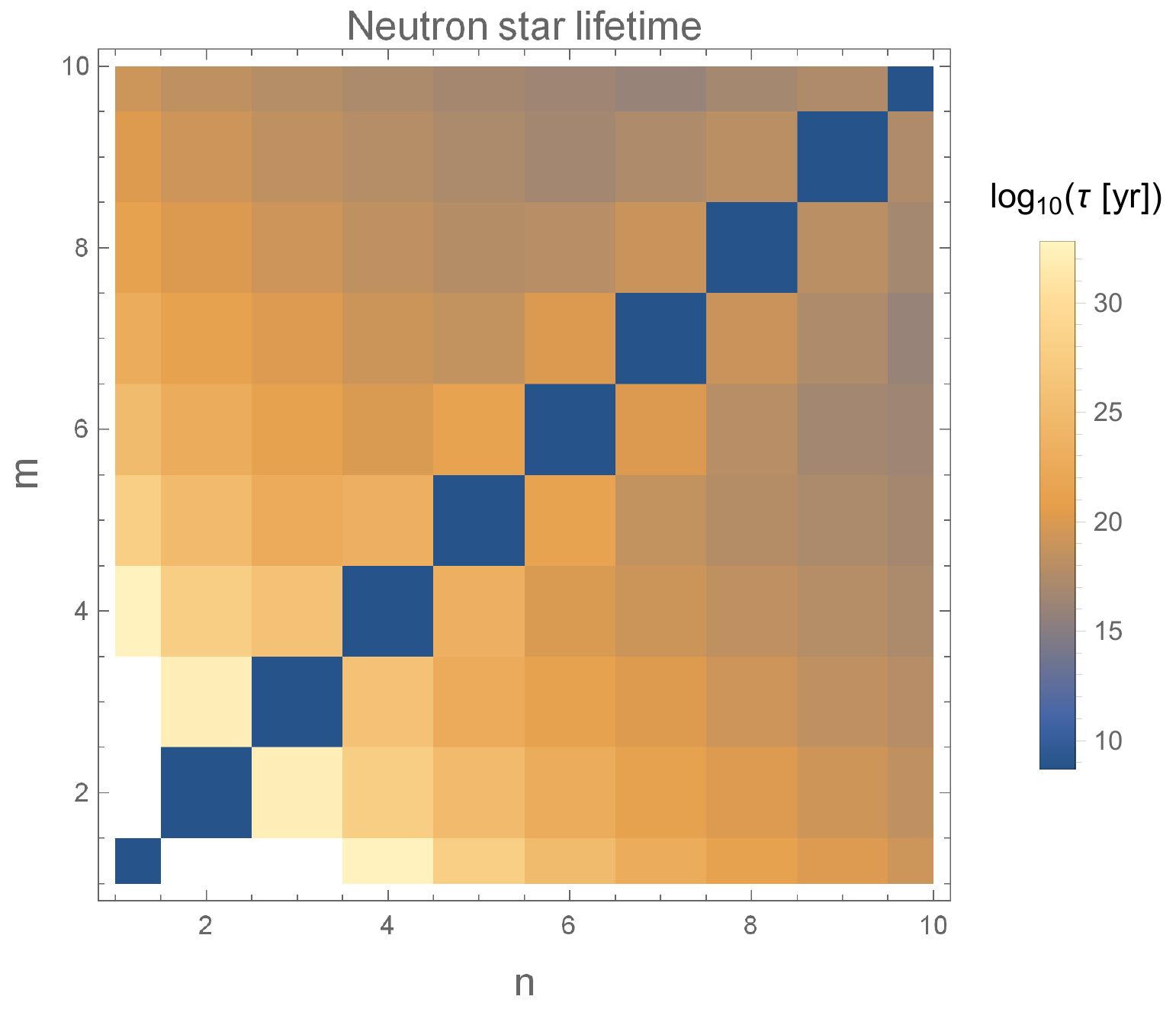}
\caption{Plot of neutron star lifetimes after being infected by a $Q$-ball with initial charge $Q_0 = 10^{25}$ as a function of various $n$, $m$ corresponding to the terms in the lifting potential. The diagonal $n=m$ is ruled out because the B-violating decays are forbidden, and the stability of the $Q$-ball causes it to grow without bound, quickly consuming the star.}
\label{fig:NSLifetimes}
\end{figure}

\begin{figure}
\centering
\includegraphics[width=0.8\linewidth]{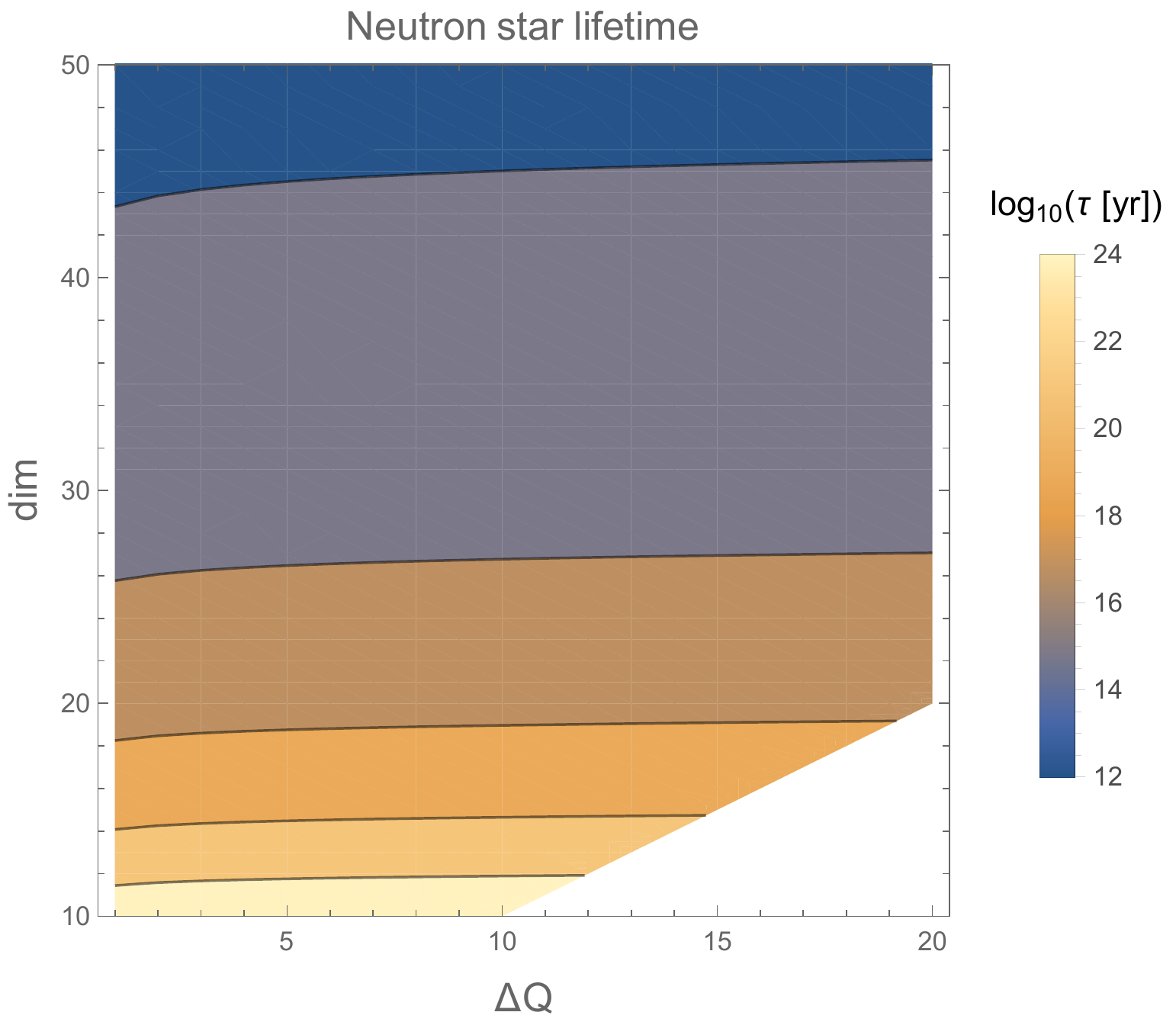}
\caption{Plot of neutron star lifetimes after being infected by a $Q$-ball with initial charge $Q_0 = 10^{25}$ as a function of the charge violation per decay $\Delta Q>0$ and the dimension of the operator in the lifting potential. The white region in the lower right corner is where $\Delta Q > \text{dim}$, which is not allowed since it implies one of either $n$ or $m$ is negative.}
\label{fig:NSLifetimesHighDim}
\end{figure}

Solving the baryon number evolution equation with $\dot{N}_0 \not= 0$ and integrating until $N_n = 0$ gives us the lifetime of an infected neutron star. This information is plotted in figure \ref{fig:NSLifetimes} and table \ref{tab:NSLifetimes}.
As we can see, the diagonal where $n=m$ is ruled out, with a lifetime of about $10^8$ years. This is due to the fact that the $Q$-ball is absolutely stable in this regime, and therefore grows without bound as it eats away at the neutron star, quickly consuming it. In fact, this is an upper limit on the lifetime; the final charge of the $Q$-ball in this situation is $3 \times 10^{57}$, which is beyond the critical charge for a flat direction $Q$-ball to change into a curved direction type, which as mentioned before, has an even higher rate of neutron consumption. The highest charge for a $Q$-ball with baryon-violating decays is only $10^{42}$, well below the critical charge. Interestingly, in the regions with operator dimension $\le 4$, the $Q$-ball decays so quickly that it breaks down completely before the neutron star is consumed. As mentioned in the previous subsection, $Q$-balls in this regime aren't stable in free space anyway. We can see that as we move away from the $n=m$ diagonal (increasing $\Delta Q$), the lifetime of the star begins to drop, then levels out, with the magnitude of the drop decreasing as the operator dimension increases. In order to study the effects of very high-dimension operators ($n+m \rightarrow \infty$), we once again use equation \ref{eq:JMC} to calculate $J_{nm}^N$ and solve the baryon number evolution equations. This is plotted in figure \ref{fig:NSLifetimesHighDim}.
What we find is quite interesting: the lifetime appears to approach a limiting value around $10^{12}$ years as the operator dimension increases. The lifetime is roughly independent of $\Delta Q$, though it does drop slightly near $\Delta Q = 0$. This appears to match the trend of figure \ref{fig:NSLifetimes} as the operator dimension is increased.

\section{Conclusion} \label{sec:Conclusion}
We have shown here that $Q$-balls can make up dark matter if baryon-violating terms of dimension $n+m > 5$ are present in the scalar potential. Cases in which there is no baryon violation ($n=m$) are ruled out as well due to unrestricted $Q$-ball growth. The baryon number violation is also necessary for the Affleck-Dine mechanism to work. This eliminates the neutron star bounds. Beyond this, there appears to be no restriction on these operators, even at very high dimension. The low level of baryon number violation does not affect the experimental limits based on IceCube \cite{Kasuya:2015uka}, Super-Kamiokande \cite{Arafune:2000yv} and other direct detection experiments.  However, one should keep in mind that $Q$-balls may carry some electric charge \cite{Kusenko:1997vp,Arafune:2000yv,Shoemaker:2008gs}, making them almost invisible to most direct-detection searches. (A positively charged $Q$-ball cannot destabilize nuclei because the Coulomb repulsion prevents any strong interactions between non-relativistic $Q$-balls and matter nuclei.)  This leaves a wide range of parameters available for dark matter in the form of supersymmetric $Q$-balls.
\newparagraph

\section{Acknowledgements}
This work was supported by the U.S. Department of Energy Grant No. DE-SC0009937.  A.K. was also supported by the World Premier International Research Center Initiative (WPI), MEXT, Japan. A.K. appreciates the hospitality of the Aspen Center for Physics, which is supported by National Science Foundation grant PHY-1066293.

\begin{widetext}

\appendix
\section{Tables of $Q$-ball and neutron star lifetimes}
These tables correspond to figures \ref{fig:QballLifetimes} and \ref{fig:NSLifetimes}, and list the lifetimes of $Q$-balls and neutron stars infected by $Q$-balls with baryon-violating decays.

\begin{center}
\begin{table}[H]
\begin{gather}
\begin{array}{|c||c|c|c|c|c|c|c|c|c|c|}
\hline
n \backslash m & 1 & 2 & 3 & 4 & 5 & 6 & 7 & 8 & 9 & 10 \\
\hline\hline
1 & \infty & 7 \times 10^{-55} & 9 \times 10^{-31} & 6 \times 10^{-6} & 1 \times 10^{20} & 4 \times 10^{45} & 7 \times 10^{71} & 1 \times 10^{97} & 2 \times 10^{123} & 2 \times 10^{149} \\
\hline
2 & 7 \times 10^{-55} & \infty & 4 \times 10^{-4} & 1 \times 10^{21} & 1 \times 10^{46} & 3 \times 10^{71} & 1 \times 10^{97} & 2 \times 10^{123} & 7 \times 10^{148} & 6 \times 10^{174} \\
\hline
3 & 9 \times 10^{-31} & 4 \times 10^{-4} & \infty & 1 \times 10^{48} & 5 \times 10^{72} & 6 \times 10^{97} & 1 \times 10^{123} & 8 \times 10^{148} & 1 \times 10^{175} & 4 \times 10^{200} \\
\hline
4 & 6 \times 10^{-6} & 1 \times 10^{21} & 1 \times 10^{48} & \infty & 9 \times 10^{99} & 3 \times 10^{124} & 4 \times 10^{149} & 1 \times 10^{175} & 5 \times 10^{200} & 9 \times 10^{226} \\
\hline
5 & 1 \times 10^{20} & 1 \times 10^{46} & 5 \times 10^{72} & 9 \times 10^{99} & \infty & 7 \times 10^{151} & 3 \times 10^{176} & 4 \times 10^{201} & 1 \times 10^{227} & 4 \times 10^{252} \\
\hline
6 & 4 \times 10^{45} & 3 \times 10^{71} & 6 \times 10^{97} & 3 \times 10^{124} & 7 \times 10^{151} & \infty & 7 \times 10^{203} & 3 \times 10^{228} & 4 \times 10^{253} & 9 \times 10^{278} \\
\hline
7 & 7 \times 10^{71} & 1 \times 10^{97} & 1 \times 10^{123} & 4 \times 10^{149} & 3 \times 10^{176} & 7 \times 10^{203} & \infty & 7 \times 10^{255} & 3 \times 10^{280} & >10^{300} \\
\hline
8 & 1 \times 10^{97} & 2 \times 10^{123} & 8 \times 10^{148} & 1 \times 10^{175} & 4 \times 10^{201} & 3 \times 10^{228} & 7 \times 10^{255} & \infty & >10^{300} & >10^{300} \\
\hline
9 & 2 \times 10^{123} & 7 \times 10^{148} & 1 \times 10^{175} & 5 \times 10^{200} & 1 \times 10^{227} & 4 \times 10^{253} & 3 \times 10^{280} & >10^{300} & \infty & >10^{300} \\
\hline
10 & 2 \times 10^{149} & 6 \times 10^{174} & 4 \times 10^{200} & 9 \times 10^{226} & 4 \times 10^{252} & 9 \times 10^{278} & >10^{300} & >10^{300} & >10^{300} & \infty \\
\hline
\end{array} \nonumber
\end{gather}
\caption{Table of $Q$-ball lifetimes (in years) for various lifting potentials. Lifetimes with an $\infty$ are absolutely stable due to restoration of the $U(1)_B$ symmetry.}
\label{tab:QballLifetimes}
\end{table}

\begin{table}[H]
\begin{gather}
\begin{array}{|c||c|c|c|c|c|c|c|c|c|c|}
\hline
n \backslash m & 1 & 2 & 3 & 4 & 5 & 6 & 7 & 8 & 9 & 10 \\
\hline\hline
1 & 5 \times 10^8 & \infty  & \infty  & 6 \times 10^{32} & 1 \times 10^{28} & 1 \times 10^{25} & 7 \times 10^{22} & 2 \times 10^{21} & 1 \times 10^{20} & 2 \times 10^{19} \\
\hline
2 & \infty  & 5 \times 10^8 & 2 \times 10^{32} & 7 \times 10^{27} & 9 \times 10^{24} & 8 \times 10^{22} & 2 \times 10^{21} & 1 \times 10^{20} & 2 \times 10^{19} & 3 \times 10^{18} \\
\hline
3 & \infty  & 2 \times 10^{32} & 5 \times 10^8 & 2 \times 10^{26} & 5 \times 10^{22} & 2 \times 10^{21} & 1 \times 10^{20} & 2 \times 10^{19} & 3 \times 10^{18} & 7 \times 10^{17} \\
\hline
4 & 6 \times 10^{32} & 7 \times 10^{27} & 2 \times 10^{26} & 5 \times 10^8 & 3 \times 10^{23} & 1 \times 10^{20} & 1 \times 10^{19} & 3 \times 10^{18} & 7 \times 10^{17} & 2 \times 10^{17} \\
\hline
5 & 1 \times 10^{28} & 9 \times 10^{24} & 5 \times 10^{22} & 3 \times 10^{23} & 5 \times 10^8 & 3 \times 10^{21} & 4 \times 10^{18} & 6 \times 10^{17} & 2 \times 10^{17} & 7 \times 10^{16} \\
\hline
6 & 1 \times 10^{25} & 8 \times 10^{22} & 2 \times 10^{21} & 1 \times 10^{20} & 3 \times 10^{21} & 5 \times 10^8 & 1 \times 10^{20} & 9 \times 10^{17} & 6 \times 10^{16} & 3 \times 10^{16} \\
\hline
7 & 7 \times 10^{22} & 2 \times 10^{21} & 1 \times 10^{20} & 1 \times 10^{19} & 4 \times 10^{18} & 1 \times 10^{20} & 5 \times 10^8 & 1 \times 10^{19} & 2 \times 10^{17} & 1 \times 10^{16} \\
\hline
8 & 2 \times 10^{21} & 1 \times 10^{20} & 2 \times 10^{19} & 3 \times 10^{18} & 6 \times 10^{17} & 9 \times 10^{17} & 1 \times 10^{19} & 5 \times 10^8 & 1 \times 10^{18} & 7 \times 10^{16} \\
\hline
9 & 1 \times 10^{20} & 2 \times 10^{19} & 3 \times 10^{18} & 7 \times 10^{17} & 2 \times 10^{17} & 6 \times 10^{16} & 2 \times 10^{17} & 1 \times 10^{18} & 5 \times 10^8 & 3 \times 10^{17} \\
\hline
10 & 1 \times 10^{19} & 3 \times 10^{18} & 7 \times 10^{17} & 2 \times 10^{17} & 7 \times 10^{16} & 3 \times 10^{16} & 1 \times 10^{16} & 7 \times 10^{16} & 3 \times 10^{17} & 5 \times 10^8 \\
\hline
\end{array} \nonumber
\end{gather}
\caption{Table of infected neutron star lifetimes (in years) for various lifting potentials. Lifetimes with an $\infty$ are absolutely stable.}
\label{tab:NSLifetimes}
\end{table}
\end{center}
\end{widetext}

\end{document}